\documentclass[11pt,twoside]{article}
\usepackage{asp2010}
\usepackage{subfigure}

\resetcounters

\begin{document}

\title{Spatio-kinematic modelling: Testing the link between planetary nebulae and close binaries}
\author{David Jones$^1$, Amy~A. Tyndall$^{2,3}$, Leo Huckvale$^2$, Barnabas Prouse$^2$ and Myfanwy Lloyd$^2$
\affil{$^1$European Southern Observatory, Alonso de C\'ordova 3107,  Santiago, Chile}
\affil{$^2$Jodrell Bank Centre for Astrophysics, University of Manchester, UK}
\affil{$^3$Isaac Newton Group of Telescopes, Santa Cruz de La Palma, Spain}
}

\begin{abstract}
It is widely believed that central star binarity plays an important role in the formation and evolution of aspherical planetary nebulae, however observational support for this hypothesis is lacking.  Here, we present the most recent results of a continuing programme to model the morphologies of all planetary nebulae known to host a close binary central star.  Initially, this programme allows us to compare the inclination of the nebular symmetry axis to that of the binary plane, testing the theoretical expectation that they will lie perpendicular - to date, all have satisfied this expectation, indicating that each nebula has been shaped by its central binary star.  

As a greater sample of nebulae are modelled, it will be possible to search for trends connecting the parameters of both nebula and central binary, strengthening our understanding of the processes at work in these objects.  I will discuss some of the more obvious comparisons, and their current statuses, as well as the obvious links to common envelope evolution.

\end{abstract}

\section{Introduction}

Planetary nebulae (PNe) are glowing shells of gas ejected by intermediate mass stars at the end of their lives.  The most widely accepted explanation for their formation is described by the Generalised Interacting Stellar Winds Theory (GISW Theory, see, e.g., review by \citealp{balick02}). 
Under this scheme, PNe are formed from the snowplough-like interaction between two stellar winds.  In order to form an aspherical PN, one of these winds is required to be aspherical.  Three main possibilities have been considered for the origin of this aspherical wind: (a) Stellar rotation, (b) Magnetic fields, and (c) Central star binarity.  Of these, only central star binarity has stood up to theoretical investigation \citep{demarco09}.  The levels of stellar rotation required to produce significant wind asphericity were found to be too high \citep{bond90}, while magnetic fields were found to be too short-lived in single star systems \citep{nordhaus07}.

\begin{table}[!ht]
\caption{A table of PNe, with well studied binary central stars, that have been the subject of detailed spatio-kinematic study.}
\label{tab:neb}
\begin{center}
{\small
\begin{tabular}{ccc}
\tableline
\noalign{\smallskip}
PN & Morphology & References\\
\noalign{\smallskip}
\tableline
\noalign{\smallskip}
Abell 41 & open-ended bipolar with equatorial torus & \citet{jones10b}\\
Abell 63 & barrel-like with bipolar ``end-caps'' & \citet{mitchell07b}\\
Abell 65 & double-shelled bipolar & \citet{huckvale10}\\
&& Huckvale et al. (in prep.)\\
HaTr~4 & Elliptical with equatorial torus & \citet{tyndall10}\\
&& Tyndall et al. (in prep.)\\
NGC~6337 & Torus without bipolar lobes, &\citet{garcia-diaz09},\\
& jet-like corkscrew outflow &  \citet{hillwig10}\\
Sp~1 & end-on bipolar & \citet{mitchell07}, \\
&&Jones et al. (in prep.)\\
\noalign{\smallskip}
\tableline
\end{tabular}
}
\end{center}
\end{table}

For binary central stars, the shaping effect is expected to be greatest in those systems that have passed through a common-envelope (CE) phase.  Here, it is the ejected CE which goes on to form the required asphericity, as it is preferentially expelled in the orbital plane of the binary \citep{Nordhaus06}.  Therefore, any PNe formed from a post-CE central binary is predicted to display a bipolar morphology, the symmetry axis of which is aligned perpendicular to the plane of the central binary.

While central star binarity remains the strongest contender for the origin of aspherical planetary nebulae, observational evidence is lacking.  In recent years, the effort to find the observational support for the so called \textit{Binary Hypothesis} has gathered pace, mainly through the search for new binary central star systems (\citealp{miszalski09a,corradi11,miszalski11,miszalski10, jones10d}; Boffin et al. these proceedings), which now number $\sim$50 (c.f.\ $\sim$3000 known Galactic PNe).  With the relatively rapid rate of discovery of new binary central star systems, it is now important to begin to try to relate the properties of the host PNe to those of their central stars.  

In these proceedings, I will detail the results of an on-going investigation to determine the morphologies of all PNe known to host binary central stars, with the aim of relating those morphologies to the parameters of the central binary stars (particularly testing the expected perpendicular alignment between nebular symmetry axis and binary plane).

\section{Determining the morphologies of planetary nebulae}

\begin{figure}
\includegraphics[width=\textwidth]{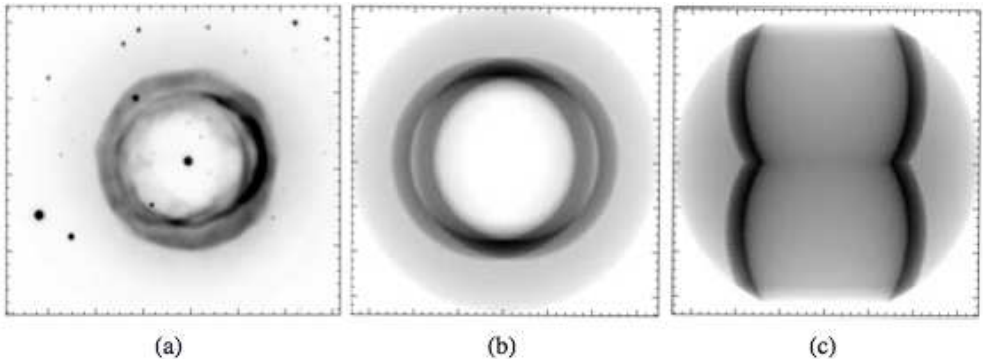}
\caption{Image and model of Sp~1. (a) Shows the observed image of Sp~1, an apparently spherical/circular PNe, (b) shows the model at the observed inclination and (c) at an inclination of 90$^\circ$ (Note: images not to same scale).  This PNe is a good example of the need for spatio-kinematic modelling due to the degeneracy between morphology and orientation in imagery.}
\label{fig:sp1}
\end{figure}

The orientation and morphology of a PN cannot be determined from imagery alone as projection effects result in a degeneracy between shape and inclination (i.e.\ a bipolar nebula viewed ``end-on'' will appear to be circular/spherical, see, e.g.\ \citealp{kwok10}).  This degeneracy can only be circumvented by spatio-kinematic modelling of the PN based on both imagery and spatially resolved spectroscopy of the nebula (see figure \ref{fig:sp1}).  Detailed explanation of the techniques used can be found in \citet{jones10a,jones10b}.

It is important to note that spatio-kinematic modelling reveals not only the general morphology of the PN, but also the orientation of the nebular structure with respect to the line of sight.  This inclination to the line of sight is an important parameter as it can be compared to the inclination of the orbital plane (of the central binary, determined by modelling of photometric and/or radial velocity measurements of the system), in order to test the expected perpendicular alignment, and therefore examine whether the PN has been shaped by its central binary.

\section{The sample}
\label{sec:PN}

\begin{figure}
\centering
\includegraphics[width=0.85\textwidth]{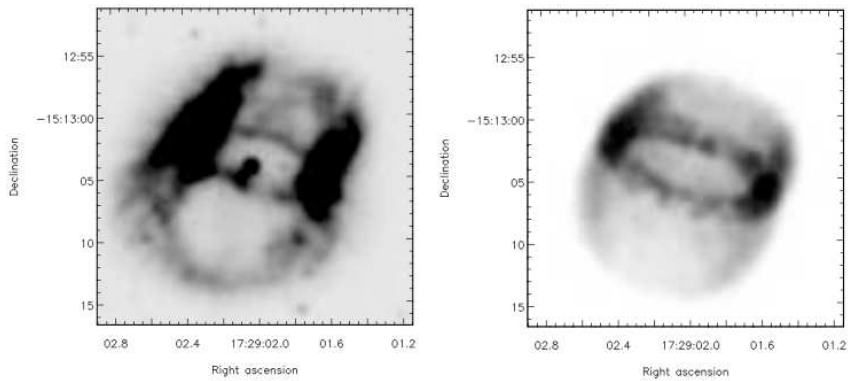}
\caption{An image of Abell 41 (left) and the \citet{jones10b} model (right).}
\label{fig:Abell 41}
\end{figure}

To date, only six PNe with well-constrained binary central stars (i.e.\ with known orbital inclinations) have been subjected to detailed spatio-kinematic study.  However, each of the six PNe has been shown to follow the expected anti-alignment between nebular symmetry axis and binary orbital plane, indicating that each of the PNe has been shaped by its central binary.  The PNe studied along with a description of the derived morphology is presented in table \ref{tab:neb}.  Examples of the observed and modelled morphologies of Sp~1 and Abell 41 are shown in figures \ref{fig:sp1} and \ref{fig:Abell 41}, respectively.
 
\begin{figure}
\centering
\includegraphics[width=0.8\textwidth]{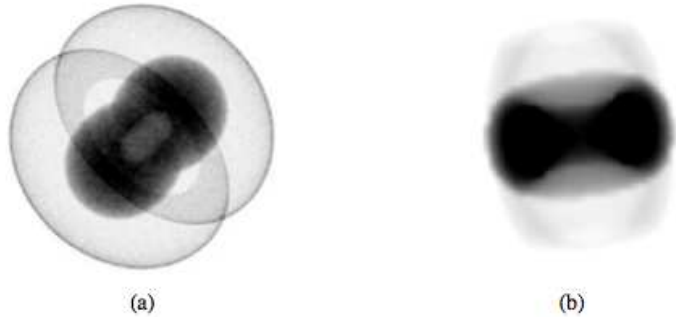}
\caption{Spatio-kinematic models of (a) Abell 65 (\citealp{huckvale10}, in preparation) and (b) HaTr~4 (\citealp{tyndall10}, in preparation).}
\label{fig:neb}
\end{figure}

\section{Results}

As detailed in section \ref{sec:PN}, all of the six PNe, which have been subjected to detailed study of both central binary system and nebular morphology, have been shown to follow the expected perpendicular alignment between nebular symmetry axis and binary orbital plane.  The fact that \emph{all} of the sample display this perpendicular alignment indicates that each PNe has been shaped in some way by its central binary - strengthening the candidacy of central star binarity as the most important factor in the shaping of aspherical PNe.  Although some common features do appear among the PNe (equatorial rings, polar outflows), it is clear that each of the PNe studied is morphologically very different (see e.g. the modelled morphologies of Abell 65 and HaTr~4, shown in figure \ref{fig:neb}), indicating that understanding the exact shaping mechanism(s) will not be trivial!

\subsection{Expected correlations?}

\begin{figure}
\centering
\includegraphics[width=\textwidth]{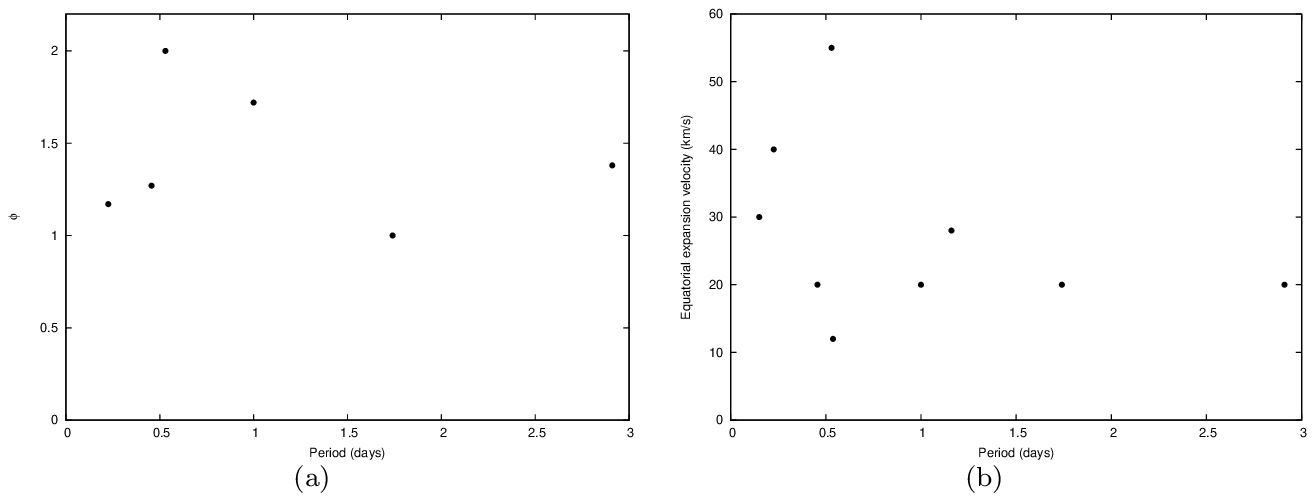}
\caption{Plots of central star orbital period versus (a) collimation factor, $\phi$, and (b) equatorial expansion velocity for the PNe in this study.}
\label{fig:plots}
\end{figure}

As the number of PN systems subjected to detailed study increases it will become possible to search for correlations between the orbital parameters of the binary central stars and the morphological parameters of their host PNe.  The most obvious correlations to examine are between the period of the central star, which should be related to the orbital angular momentum deposited in the ejected CE, and the nebular morpho-kinematics.  Examining the relationship between binary period and degree of collimation of the nebular lobes (characterised by a factor I have called $\phi$, which is the ratio of maximum lobe radius and waist radius), one might expect the most highly collimated PNe (high $\phi$) to host the shortest period binaries (as more orbital angular momentum deposited in the CE will cause a high over-density in the binary plane, creating the tight waist of the collimated PN).  Orbital period is plotted against the collimation in figure \ref{fig:plots}a (ETHOS~1, a seventh PNe for which the relevant data is available, has also been included - see \citealp{miszalski11}), showing no strong correlation.  However, the most collimated PNe do exhibit the lowest orbital periods, indicating that a correlation may be found with the addition of further data points.

If, as current theories predict, the ejected CE goes on to form the waist of the subsequent PNe, one might expect to find a correlation between the equatorial expansion velocity of the PNe and the orbital periods of their central binary stars (as lower periods require more angular momentum to be transferred to the CE, possibly ejecting it at higher velocity).  These two parameters are plotted in figure \ref{fig:plots}b (including a further 3 PNe for which the relevant data is available - ETHOS~1, The Necklace, \& NGC~6778 - see \citealp{miszalski11}, \citealp{corradi11} and \citealp{maestro04}, respectively), again showing no apparent correlation.  However, the greatest equatorial expansion velocities are found in PNe with the shorted orbital period central stars, hinting that a possible correlation may be found with the addition of further data points.

\subsection{Pre-CE polar outflows?}
\label{sec:polar}

\begin{table}[!ht]
\caption{The kinematical ages of PNe, with binary central stars, and their observed polar outflows (see text for references).}
\label{tab:jets}
\begin{center}
{\small
\begin{tabular}{ccc}
\tableline
\noalign{\smallskip}
PN & PN age & Polar outflow age\\
& (yrs kpc$^{-1}$) & (yrs kpc$^{-1}$)\\
\noalign{\smallskip}
\tableline
\noalign{\smallskip}
Abell 63 & $3500\pm200$ & $5200\pm1200$ \\
ETHOS~1 & $900\pm100$ & $1750\pm250$\\
The Necklace & $1100\pm100$ & $2350\pm450$ \\
NGC~6337 & $\sim9200$ & $\sim1000$ \\
\noalign{\smallskip}
\tableline
\end{tabular}
}
\end{center}
\end{table}

ETHOS~1 \citep{miszalski11}, The Necklace \citep{corradi11}, NGC~6337 \citep{garcia-diaz09} and Abell 63 \citep{mitchell07b} all display extended polar outflows in the form of jets or end-caps, which have been kinematically studied.  Interestingly, in three out of the four systems (ETHOS~1, The Necklace \& Abell 63) the polar outflows are found to be older than the central PN.  Currently, the only explanation for this discrepancy is for the polar outflows to have formed \emph{before} the central binary entered the CE phase, possibly at the onset of mass transfer.  Furthermore, in each of these systems the polar outflows are found to be approximately $\sim$1000 years older than the central PNe (see table \ref{tab:jets}), which is consistent with the expected lifetime of the CE phase \citep{webbink08}.

\section{Discussion and conclusions}

Each of the PNe studied shows the expected perpendicular alignment between PN and binary orbital plane, indicating that central star binarity has played an important role in the shaping of the PNe.  These results indicate that central star binarity may indeed be the most important factor in the formation of aspherical PNe - a key unsolved problem in the PN community.

Too few PNe have been subjected to detailed study to be able to determine strong correlations between the parameters of nebula and central binary, however the results so far are certainly encouraging enough to warrant continued investigation.

Probably the most striking result to come from this work, particularly in a binary context, is that polar outflows in PNe may originate from before the CE phase, meaning that PNe may be used as a tool to constrain the mass-loss leading up to the CE phase \emph{as well as} the ejection of the CE itself.


\bibliography{djones.bib}

\begin{thebibliography}{}
\expandafter\ifx\csname natexlab\endcsname\relax\def\natexlab#1{#1}\fi
\expandafter\ifx\csname url\endcsname\relax
  \def\url#1{\texttt{#1}}\fi
\expandafter\ifx\csname urlprefix\endcsname\relax\def\urlprefix{URL }\fi
\providecommand{\eprint}[2][]{\url{#2}}

\bibitem[{Balick \& Frank(2002)}]{balick02}
Balick, B., \& Frank, A. 2002, ARA\&A, 40, 439

\bibitem[{Bond \& Livio(1990)}]{bond90}
Bond, H.~E., \& Livio, M. 1990, ApJ, 335, 568

\bibitem[{{Corradi} et~al.(2011){Corradi}, {Sabin}, {Miszalski},
  {Rodr{\'{\i}}guez-Gil}, {Santander-Garc{\'{\i}}a}, {Jones}, {Drew},
  {Mampaso}, {Barlow}, {Rubio-D{\'{\i}}ez}, {Casares}, {Viironen}, {Frew},
  {Giammanco}, {Greimel}, \& {Sale}}]{corradi11}
{Corradi}, R.~L.~M., {Sabin}, L., {Miszalski}, B., {Rodr{\'{\i}}guez-Gil}, P.,
  {Santander-Garc{\'{\i}}a}, M., {Jones}, D., {Drew}, J.~E., {Mampaso}, A.,
  {Barlow}, M.~J., {Rubio-D{\'{\i}}ez}, M.~M., {Casares}, J., {Viironen}, K.,
  {Frew}, D.~J., {Giammanco}, C., {Greimel}, R., \& {Sale}, S.~E. 2011, MNRAS,
  410, 1349

\bibitem[{{de Marco}(2009)}]{demarco09}
{de Marco}, O. 2009, PASP, 121, 316

\bibitem[{{Garc{\'{\i}}a-D{\'{\i}}az} et~al.(2009){Garc{\'{\i}}a-D{\'{\i}}az},
  {Clark}, {L{\'o}pez}, {Steffen}, \& {Richer}}]{garcia-diaz09}
{Garc{\'{\i}}a-D{\'{\i}}az}, M.~T., {Clark}, D.~M., {L{\'o}pez}, J.~A.,
  {Steffen}, W., \& {Richer}, M.~G. 2009, ApJ, 699, 1633

\bibitem[{{Hillwig} et~al.(2010){Hillwig}, {Bond}, {Af{\c s}ar}, \& {De
  Marco}}]{hillwig10}
{Hillwig}, T.~C., {Bond}, H.~E., {Af{\c s}ar}, M., \& {De Marco}, O. 2010, AJ,
  140, 319

\bibitem[{{Huckvale} et~al.(2011){Huckvale}, {Prouse}, {Jones}, {Lloyd},
  {O'Brien}, \& {Pollacco}}]{huckvale10}
{Huckvale}, L., {Prouse}, B., {Jones}, D., {Lloyd}, M., {O'Brien}, T.~J., \&
  {Pollacco}, D. 2011, in {Asymmetric Planetary Nebulae 5}, edited by
  {{Zijlstra}, A.~A. and {McDonald}, I. and {Lagadec}, E.}, {Asymmetric
  Planetary Nebulae}, 119

\bibitem[{{Jones} et~al.(2010{\natexlab{a}}){Jones}, {Lloyd}, {Mitchell},
  {Pollacco}, {O'Brien}, \& {Vaytet}}]{jones10a}
{Jones}, D., {Lloyd}, M., {Mitchell}, D.~L., {Pollacco}, D.~L., {O'Brien},
  T.~J., \& {Vaytet}, N.~M.~H. 2010{\natexlab{a}}, MNRAS, 401, 405

\bibitem[{{Jones} et~al.(2010{\natexlab{b}}){Jones}, {Lloyd},
  {Santander-Garc{\'{\i}}a}, {L{\'o}pez}, {Meaburn}, {Mitchell}, {O'Brien},
  {Pollacco}, {Rubio-D{\'{\i}}ez}, \& {Vaytet}}]{jones10b}
{Jones}, D., {Lloyd}, M., {Santander-Garc{\'{\i}}a}, M., {L{\'o}pez}, J.~A.,
  {Meaburn}, J., {Mitchell}, D.~L., {O'Brien}, T.~J., {Pollacco}, D.,
  {Rubio-D{\'{\i}}ez}, M.~M., \& {Vaytet}, N.~M.~H. 2010{\natexlab{b}}, MNRAS,
  408, 2312

\bibitem[{{Jones} et~al.(2011){Jones}, {Pollacco}, {Faedi}, \&
  {Lloyd}}]{jones10d}
{Jones}, D., {Pollacco}, D., {Faedi}, F., \& {Lloyd}, M. 2011, in {Asymmetric
  Planetary Nebulae 5}, edited by {{Zijlstra}, A.~A. and {McDonald}, I. and
  {Lagadec}, E.}, {Asymmetric Planetary Nebulae}, 117

\bibitem[{{Kwok}(2010)}]{kwok10}
{Kwok}, S. 2010, PASA, 27, 174

\bibitem[{{Maestro} et~al.(2004){Maestro}, {Guerrero}, \&
  {Miranda}}]{maestro04}
{Maestro}, V., {Guerrero}, M.~A., \& {Miranda}, L.~F. 2004, in Asymmetrical
  Planetary Nebulae III: Winds, Structure and the Thunderbird, edited by
  {M.~Meixner, J.~H.~Kastner, B.~Balick, \& N.~Soker}, vol. 313 of Astronomical
  Society of the Pacific Conference Series, 127

\bibitem[{{Miszalski} et~al.(2009){Miszalski}, {Acker}, {Moffat}, {Parker}, \&
  {Udalski}}]{miszalski09a}
{Miszalski}, B., {Acker}, A., {Moffat}, A.~F.~J., {Parker}, Q.~A., \&
  {Udalski}, A. 2009, A\&A, 496, 813

\bibitem[{{Miszalski} et~al.(2011{\natexlab{a}}){Miszalski}, {Corradi},
  {Boffin}, {Jones}, {Sabin}, {Santander-Garc{\'{\i}}a},
  {Rodr{\'{\i}}guez-Gil}, \& {Rubio-D{\'{\i}}ez}}]{miszalski11}
{Miszalski}, B., {Corradi}, R.~L.~M., {Boffin}, H.~M.~J., {Jones}, D., {Sabin},
  L., {Santander-Garc{\'{\i}}a}, M., {Rodr{\'{\i}}guez-Gil}, P., \&
  {Rubio-D{\'{\i}}ez}, M.~M. 2011{\natexlab{a}}, MNRAS, 413, 1264

\bibitem[{{Miszalski} et~al.(2011{\natexlab{b}}){Miszalski}, {Corradi},
  {Jones}, {Santander-Garc\'ia}, {Rodr\'iguez-Gil}, \&
  {Rubio-D\'iez}}]{miszalski10}
{Miszalski}, B., {Corradi}, R.~L.~M., {Jones}, D., {Santander-Garc\'ia}, M.,
  {Rodr\'iguez-Gil}, P., \& {Rubio-D\'iez}, M.~M. 2011{\natexlab{b}}, in
  {Asymmetric Planetary Nebulae 5}, edited by {{Zijlstra}, A.~A. and
  {McDonald}, I. and {Lagadec}, E.}, {Asymmetric Planetary Nebulae}, 328

\bibitem[{Mitchell(2007)}]{mitchell07}
Mitchell, D.~L. 2007, Ph.D. thesis, University of {M}anchester

\bibitem[{Mitchell et~al.(2007)Mitchell, Pollacco, O'Brien, Bryce, Lopez,
  Meaburn, \& Vaytet}]{mitchell07b}
Mitchell, D.~L., Pollacco, D., O'Brien, T.~J., Bryce, M., Lopez, J.~A.,
  Meaburn, J., \& Vaytet, N.~M.~H. 2007, MNRAS, 374, 1404

\bibitem[{Nordhaus \& Blackman(2006)}]{Nordhaus06}
Nordhaus, J., \& Blackman, E.~G. 2006, MNRAS, 370, 2004

\bibitem[{{Nordhaus} et~al.(2007){Nordhaus}, {Blackman}, \&
  {Frank}}]{nordhaus07}
{Nordhaus}, J., {Blackman}, E.~G., \& {Frank}, A. 2007, MNRAS, 376, 599

\bibitem[{{Tyndall} et~al.(2011){Tyndall}, {Jones}, {Lloyd}, {O'Brien},
  {Pollacco}, \& {Mitchell}}]{tyndall10}
{Tyndall}, A., {Jones}, D., {Lloyd}, M., {O'Brien}, T.~J., {Pollacco}, D.~L.,
  \& {Mitchell}, D.~L. 2011, in {Asymmetric Planetary Nebulae 5}, edited by
  {{Zijlstra}, A.~A. and {McDonald}, I. and {Lagadec}, E.}, {Asymmetric
  Planetary Nebulae}, 121

\bibitem[{{Webbink}(2008)}]{webbink08}
{Webbink}, R.~F. 2008, in Astrophysics and Space Science Library, edited by
  {E.~F.~Milone, D.~A.~Leahy, \& D.~W.~Hobill}, vol. 352 of Astrophysics and
  Space Science Library, 233

\end{thebibliography}

\end{document}